\documentclass[12pt,a4papper]{article}
\usepackage{amsfonts,amsmath,amssymb}
\usepackage{mathrsfs}
\usepackage{amscd}

\def\({\left(}
\def\){\right)}
\def\[{\left[}
\def\]{\right]}

\def\a{\alpha}
\def\be{\beta}
\def\g{\gamma}

\def\de{\delta}

\def\ve{\varepsilon}
\def\la{\lambda}

\def\vf{\varphi}
\def\s{\sigma}

\def\build#1_#2^#3{\mathrel{\mathop{\kern
																			0pt#1}\limits_{#2}^{#3}}}

\def\beq{\begin{equation}}
\def\eeq{\end{equation}}

\def\beqa*{\begin{eqnarray*}}
\def\eeqa*{\end{eqnarray*}}

\def\ba{\hspace{-8mm}\begin{array}{lll}}
\def\ea{\end{array}}

\def\bal*{\begin{align*}}

\def\enal*{\end{align*}}

\newcommand\VEC[1]{\boldsymbol{#1}}

\begin{document}

\title{The role of translational invariance in non linear gauge theories of gravity}

\author{J. Mart\'in\thanks{Departamento de F\'{\i}sica
Fundamental, Universidad de Salamanca, 37008 Salamanca, Spain,
e-mail: chmm@usal.es} \and A. Tiemblo\thanks{Instituto de Matem\'aticas y
F\'{\i}sica Fundamental, CSIC, Serrano 113b, 28006 Madrid, Spain,
e-mail: tiemblo@imaff.cfmac.csic.es}}
\date{}

\maketitle

\begin{abstract}

The internal structure of the tetrads in a Poincar\'e non linear gauge theory of gravity is considered. Minkowskian coordinates becomes dynamical degrees of freedom playing the role of Goldstone bosons of the translations. A critical length allowing a covariant expansion similar to the weak field approach is deduced, the zeroth order metric being maximally symmetric (Minkowskian in some cases).

\end{abstract}

\newpage
\section{Introduction}

General Relativity was initially proposed as a geometrical theory constructed on the general background of the Differential Geometry. Nevertheless the coupling between half--integer spin and gravity puts in evidence the necessity to enlarge the geometrical framework with the inclusion of a local internal group having fermionic representation as required by the universal character of the gravitational field. At the same time the possibility to treat gravity on the same way that the remaining forces appears. In the search for the unification of all the interactions, many different alternatives to Einstein's geometrical point of view have been suggested \cite{Uti56}--\cite{Ash86}. As a consequence the close analogy between gauge theories of gravity and the usual Yang--Mills approaches to the standard model becomes more and more evident, which allows us to think about gravity on the same lines as for the remaining forces.

However there exists in the case of the gravity a very characteristic object without a clear partner in the description of any other interaction. As a matter of fact the tetrads are something like the footprint of geometry, that, in this way, remains still present in these schemes. Thus, to reach a better understanding, a deeper insight on the nature of the tetrads in terms of the dynamical theories of local space--time symmetries is relevant. The local treatment of space--time groups follows, in general, the same lines of the ordinary Yang--Mills theories. Nevertheless they include, as a natural feature, the presence of translations whose meaning as gauge objects is not so clear as it happens, for instance, for spin rotations. Its role as a cornerstone in the description of gravity has pointed out in several papers \cite{Hehl74, Hehl76, Gro95, Hehl95}.

Among the different theoretical points of view, we claim that the key point to interpret the nature of the tetrads as a gauge field is precisely the treatment of the translations in a local space--time symmetry, and this is the main purpose of this paper. With this aim we are going to consider, as the most simple and natural choice, the Poincar\'e group as fundamental space--time symmetry, whose role in the description of well established conservation laws is out of any discussion. We recall however that the different matter fields are in fact spin representations, but with given values of the momenta, being this fact essential in the manner in which matter occurs in the nature. As we are going to see, to take this issue item into account the non linear realizations of groups reveals as the natural tool to describe the role of translations in a local theory. Not only but some relevant features of the gravitational field as, for instance, the weak field approximation appear, not as a phenomenological hypothesis, but as a theoretical consequence in a very natural manner. Accordingly we are going to start with a brief summary abstract of the non linear realization of groups, including the corresponding gauge local treatment. Then we will continue with the application to the translations as contained in the Poincar\'e group, which provides us with the structure and dynamical behavior of tetrads when included in the gravity gauge action.

\section{Non linear realizations}

The non linear realizations of symmetry groups is a mathematical method which was firstly introduced in field theories to establish the structure and properties of phenomenological lagrangians, and more recently in the study of space--time symmetries \cite{Cole69}--\cite{Tresg00}.

Let be $G=\{g\}$ a Lie group having a suitable subgroup $H=\{h\}$, whose linear representations $\rho(h)$ are known, acting on functions $\psi$ belonging to a linear space. The elements of the quotient space $G/H$ are equivalence classes of the form $gH$ (left) that define a complete partition of the group space. We call the elements of the quotient space \emph{cosets to the left (right)} with respect to $H$. Since we deal with Lie groups, the elements of $G/H$ are labelled by continuous parameters say $\xi$. We represent the elements of $G/H$ by means of these coset indicators $C(\xi)$ parametrized by the coset parameter $\xi$, playing the role of a certain kind of coordinates (see Appendix A). The non linear coset realizations expresses the action of the group $G$ on the quotient space $G/H$, i.e., on a partition of its own space. An arbitrary element $g\in G$ acts on $G/H$ transforming one coset into another
\begin{align*}
g:\  &G/H \to G/H
\\[1ex]
&C(\xi) \to C(\xi')
\end{align*}
according to the group law
\beq
\label{1}
g\, C(\xi) = C(\xi')\, h(\xi,g)
\eeq
Obviously the elements $h(\xi,g)$ appearing in the last belong to the subgroup $H$, which is going to play the role of a criteria to identify the representations of $H$, which we will call in the following the classification subgroup, since the elements $g$ of the whole group $G$ considered in \eqref{1} act non linearly on the representation space of the classification subgroup $H$ according to
\beq
\label{2}
\psi' = \rho\big[h(\xi,g)\big] \psi(z)
\eeq
where $\rho$, as mentioned above, is a linear representation of $H$ in the space of the fields $\psi$. Therefore, the action of the total group projects on the representation of the subgroup $H$ through the dependence of $h(\xi,g)$ on the group element $g$, as given by \eqref{2}. Notice that the action of the group is given on the pairs $(\xi,\psi)$. Of course it reduces to the ordinary linear action of the $H$ when we take in particular for $g$ an element of $H$. In a few words non linear realizations is a technique to realized a group on the representation of a subgroup.

To enlarge this scheme to the framework of gauge symmetries we must define a covariant differential transforming like \eqref{2}, under local transformations, thus we need a suitable non linear concept of connection. We define it in the form
\beq
\label{3}
\Gamma = C^{-1} D C
\eeq
where the covariant differential on the coset space is defined as
\beq
\label{4}
DC = (d + \Omega) C
\eeq
$\Omega$ being the ordinary linear connection of the whole group transforming as
$$
\Omega' = g \Omega g^{-1} + g dg^{-1}\,.
$$
It is very easy to verify that the non linear gauge field $\Gamma$ defined in \eqref{3} transforms as
\beq
\label{5}
\Gamma' = h \Gamma h^{-1} + h dh^{-1}
\eeq
thus allowing us to define finally the non linear covariant differential operator
\beq
\label{6}
D = d + \Gamma\,.
\eeq
It can be seen from \eqref{5} that only the components of $\Gamma$ related to the generators of $H$ behave as true connections, transforming inhomogeneously, whereas the components of $\Gamma$ over the generators associated with the cosets transform as tensors with respect to the subgroup $H$. In this last case we deal with gauge objects that can be interpreted as non linear connections or probably better as covariant differentials of the coset parameters $\xi$ that acquire, as we are going to see, the nature of Goldstone bosons.We recall that, as we have mentioned above, the coset parameters play the role of a kind of, let us say, internal coordinates depending nevertheless to the space--time coordinates introduced to define the local action of the group.

To complete this Section we notice that combining \eqref{3} and \eqref{4} one immediately obtains the relation
\beq
\label{7}
\Gamma = C^{-1}(\xi)(d+\Omega) C(\xi)
\eeq
which gives the value of the non linear connection $\Gamma$ in terms of the linear one $\Omega$. A similar expression can be formally deduced to relate non linear representations $\psi^{NL}$ with the linear one $\psi^L$, namely,
\beq
\label{8}
\psi^{NL} =  C^{-1}(\xi) \psi^L
\eeq
This last can be interpreted as describing a non linear state as a linear one dressed, through $C^{-1}(\xi)$, with the coset parameters. A point of view which is going to be exploited in the following.

\section{Non linear realization of the Poincar\'e group}

The best candidate for  a gauge theory of gravity lead us to the Poincar\'e group as the most simple and probably realistic one. Thus we are going to apply the non linear approach to this case to find out the dynamical content of such a theory. Diverse non linear realizations are possible depending on the election of the classification subgroup $H\!\subset\! G$ adopted. In this Section we take $H$ to be the Lorentz group, yielding an explicitly covariant four--dimensional formalism which provides the geometrical basis for a lagrangian approach.

Our starting point is the fundamental transformation law \eqref{1} of non linear realizations. After rewriting for $G\!=$ Poincar\'e and $H\!=$ Lorentz we parametrize the infinitesimal Poincar\'e group element $g\!\in\! G$ and the infinitesimal Lorentz elements $h\in H$ respectively as:
\begin{align}
&g= e^{i\ve^d \VEC{P}_d} e^{i\be^{ab}\VEC{J}_{ab}} \approx \VEC{I} + i(\ve^d \VEC{P}_d + \be^{ab}\VEC{J}_{ab}) \label{9}
\\[1.5ex]
&h=  e^{i\mu^{ab}\VEC{J}_{ab}} \approx \VEC{I} + i \mu^{ab}\VEC{J}_{ab} \label{10}
\end{align}
where $\VEC{J}_{ab}$ are the Lorentz generators and $\VEC{P_d}$ the translational ones. As read out from the previous discussion the left action on the elements
$$
C(\vf) = e^{i\vf^d \VEC{P}_d}
$$
of the coset space $G/H$, being the last identical with the elements of the translation group labelled by the finite parameters $\vf^d$, induces a right action of the form
$$
C'(\vf) = e^{i(\vf^d + \de\vf^d) \VEC{P}_d}
$$
Using \eqref{9} and \eqref{10} and taking the commutations relations of the Poincar\'e group algebra
\begin{align*}
&\big[\VEC{P}_a, \VEC{P}_b\big] = 0
\\[1ex]
&\big[\VEC{J}_{ab}, \VEC{P}_d\big] = i\big(\eta_{da}\VEC{P}_b - \eta_{db}\VEC{P}_a\big)
\\[1ex]
&\big[\VEC{J}_{ab}, \VEC{J}_{cd}\big] = i\big(\eta_{ac}\VEC{J}_{bd} - \eta_{ad}\VEC{J}_{bc} + \eta_{bd}\VEC{J}_{ac}
- \eta_{bc}\VEC{J}_{ad}\big) 
\end{align*}
with $\eta_{ab} = diag(-1,+1,+1,+1)$.

A simple calculation with the help of the Hausdorff--Campbell formula (see Appendix B) yields to the value of $\mu^{ab}$ in \eqref{10} as much as the variation of the translational coset parameters, namely
\beq\label{11}
\mu^{ab} = \be^{ab}\ ,\quad \de\vf^a = -\be^a_d \vf^d 
- \ve^a
\eeq
showing that $\vf^a$ transforms exactly as Minkowskian coordinates.

Being a local theory, attention must be paid to the fact that $\vf^a(x)$ behaves like an internal degrees of freedom or, in other words, ``internal" Minkowskian coordinates depending on the choice of the space--time coordinates $\{x^\a\}$.

On the other hand, using for the linear connection of the Poincar\'e group the notation
$$
\Omega = -i \Gamma^d\VEC{P}_d -i A^{ab}\VEC{J}_{ab}
$$
whose components on the Poincar\'e algebra are the  linear translational contribution $\Gamma_{_T}^d$ and the Lorentz one $A^{ab}$ respectively, we find the non linear connection to be
$$
\Gamma = -i e^d \VEC{P}_d -i A^{ab}\VEC{J}_{ab}
$$
with the Lorentz connection unmodified with respect to the linear case, but with the translational one transformed into
\beq\label{12}
e^d = D\vf^d + \Gamma_{_T}^d
\eeq
being $D\vf^d$ the formal differential covariant with respect to Lorentz group. A direct computation leads us to the transformation law
\beq\label{13}
\de e^a = -e^d \be_d^{\ a}
\eeq
and
\beq\label{14}
\de A^{ab} = D\be^{ab}
\eeq
here $D\be^{ab}$ is, as in \eqref{12}, the formal covariant Lorentz differential as requested in a local theory.

The relevant result is condensed in equation \eqref{12}. According to it, instead of the linear translational connection $\Gamma_{_T}^a$ transforming inhomogeneously as
$$
\de\Gamma_{_T}^a = - \Gamma_{_T}^d \be_d^{\ a} + D\ve^a
$$
we obtain a non linear translational connection 1--form \eqref{12} which is Lorentz covector valued. The latter will be consequently identified  from now as the Lorentz coframe or tetrad. We emphasized that this achievement allows us to interpret the tetrad as a pure gauge object, a non linear connection or, as mentioned above, as the covariant differential of an ``internal" set of Minkowskian coordinates $\{\vf^a\}$ that, as deduced from \eqref{11}, behaves as Goldstone bosons associated to the translational group. We recall that this implies to give a dynamical interpretation to the  MInkowskian coordinates.

With this assumptions one has an interpretation of the nature of the tetrads in the gauge approach to gravity. Nevertheless equation \eqref{12} strongly suggest the possibility to reach a deeper insight on the dynamical meaning of the gravitational variables. Let us notice, in first place, the atypical dimensionality of the translational connection $ \Gamma_{_T}$ in equation \eqref{12}. Being a connection an object with the same formal dimensionality of a derivative, we introduce a characteristic length,  say $\la$, to render it homogeneous with respect to the ordinary Lorentz connection $A^{ab}$. Thus we rewrite explicitly the tetrad in the form
\beq\label{15}
e^b_\a = e(0)^b_\a + \la\Gamma_{{_T}\a}^b
\eeq
with
$$
e(0)^b_\a = D_\a\vf^b = \partial_\a\vf^b + A_\a^{bc}\vf_c
$$
here latin indices refer, as above, to Lorentz group and the greek ones describe general space--time coordinates.

Being the auxiliary tetrad $e(0)^b_\a$ a covariant derivative it constitutes the ``minimal" structure suitable to change holonomic and non holonomic objects between them. Thus \eqref{15} admits an alternative writing of the form
\beq\label{16}
e^b_\a = e(0)^c_\a\big(\de^b_c + \la\Gamma_{{_T}c}^b\big)
\eeq
Now the question which naturally arises is the possibility to understand \eqref{16} as the starting point for a kind of weak field approximation for the tetrad. This naturally implies the assumption that the parameter $\la$ can be considered as a small perturbation of the auxiliary tetrad $e(0)^b_\a$.

The existence of a critically minimal length is not a new idea. A fundamental length, for instance, appears at the Planck scale \cite{Borz88,Gar95}, lattice quantum gravity has been also suggested \cite{Ber93,Fei84}, string theories also contains a minimal length \cite{Kat90,Kon90}. Of course we are not here dealing with a quantum approach. Thus the smallness of $\la$ is here introduced as a phenomenological approach and this formalism is a framework extremely well adapted to introduce this idea in a very natural way, specially when space-time translations in the Minkowskian--like internal coordinates are concerned, as it is the case.

As we will see the smallness of $\la$ leads us to a coherent result, that can be, at the same time, a theoretical support for the treatment that considers a good approximation to take gravity as a perturbation of a Minkowskian metric.

The main question now is evidently to investigate the nature and properties of the ``minimal"  geometrical background associated to the auxiliary tetrad $e(0)^b_\a$, a question  to be answered from the gravitational dynamics itself.

\section{Minimal gravity}

We are dealing with the assumption that gravity can be described as a small perturbation of a minimal structure $e(0)_\nu^a$ characterized by a critical length $\la$. Thus 
the main question is to find out the properties and geometrical meaning of this background that can be in some sense considered as a minimal gravity.

In this aim we start in this Section from the gauge action of gravity given in the form
\beq\label{17}
S = \int d^4x\,e\big(e^{\mu a}e^{\nu b}F_{\mu\nu ab} + O\big)
\eeq
Here $e^{\mu a}$ is the contravariant tetrad associated to \eqref{16}, verifying
$$
\eta^{ab}e^{\mu_a }e_{\nu b} = \de^\mu_\nu\  ,\quad e={\rm det} e^a_\mu
$$
and
$$
F_{\mu\nu ab}  = \partial_{[\mu}A_{\nu]ab} + A_{[\mu ac}A_{\nu] \ b}^{\ \ c}
$$
is the field strength tensor constructed with the Lorentz connection $A_{\mu ab}$. A term $O$ is introduced to include other possible contributions to take into account step by step as, for instance, an energy--momentum source or eventually Lagrange multipliers. Varying formally in $e_\mu^a$ and $A_{\mu ab}$ one gets
\begin{align}
\de S&= \int d^4 x\,\Big\{\left[e e^{\nu a}(F\eta_{ab}-2 F_{\mu a}e^\mu_b)\right]\de e^b_\nu
 \notag
\\[1.ex]
&\quad+(-\partial_\mu M^{\mu a\nu b}+ M^{\nu a \mu c}A_{\mu \ c}^{\  b} + M^{\mu c \nu b}A_{\mu c}^{\  \ a})\de A_{\nu ab} + \de(e O)\Big\}
\label{18}
\end{align}
where for the sake of simplicity we have introduce the following notations
\begin{align*}
&e^{\mu a} e^{\nu b}F_{\mu\nu ab} \equiv F
\\[1ex]
& e^{\nu b}F_{\mu\nu ab} \equiv F_{\mu a}
\\[1ex]
&e\,e^{\mu [a} e^{\nu b]} \equiv M^{\mu a\nu b} 
\end{align*}
When $e_\nu^a$ and $A_{\nu ab}$ are taken as the dynamical degrees of freedom, the Eins\-tein's field equations are readily obtained. Nevertheless we are interested in the internal structure of the tetrad as given by equation \eqref{16}. Thus we must explicitly introduce in $\de e^b_\nu$ the single dynamical contributions of $\vf^a$, $A_{\mu ab}$ and $\Gamma^a_{_{T}b}$.

The calculation is highly simplified with the addition in the action of a Lagrange multiplier term of the form
$$
e\,\Delta^\mu_a\big[e(0)_\mu^a -\partial_\mu\vf^a- A_{\mu\ b}^{\ a}\vf^b\big]\,.
$$
Since the definition of the minimal tetrad $e(0)_\mu^a$ is explicitly considered when we vary with respect to the multiplier $\Delta^\mu_a$ we can take then $e(0)_\mu^a$, $\Gamma^{T a}_b$, $\vf^a$ and $A_{\mu ab}$ as the dynamical variables.

With these assumptions and after a little algebra the field equations becomes
\begin{align}
&e(0)^a_\mu = D_\mu\vf^a \equiv \partial_\nu\vf^a + A_{\mu\ b}^{\ a}\vf^b \label{19}
\\[1ex]
& e^{\nu a}\big(F \eta _{ab} - 2 F_{\mu a} e^\mu_b\big) + 2 \chi\, T^\nu_b = 0
\label{20}
\\[1ex]
&\partial_\mu M^{\mu a\nu b} - M^{\nu a\mu c} A_{\mu\ c}^{\ b} - M^{\mu c\nu b} A_{\mu c}^{\ \ a}\label{21} = 0
\end{align}
where the term $2\chi\,T^\nu_b$ ($\chi=$ constant) appearing in \eqref{20} takes into account the possible presence of an energy--momentum tensor source.

To get closer to the geometrical language we adopt for $A_{\mu ab}$ the familiar decomposition
\beq\label{22}
A_{\mu ab} = e^\a_a\nabla_\mu e_{\alpha b} +\bar A_{\mu ab}
\eeq
where $\nabla_\mu$ stands for the ordinary Christoffel covariant derivative with respect to the coordinate indexes with a metric tensor defined as $g_{\mu\nu} = \eta^{ab}e_{\mu a}e_{\nu b}$. The first term in \eqref{22} giving the metric compatible part of the connection and $\bar A_{\mu ab}$ being a torsion contribution.

As it is well known decomposition \eqref{22} is the usual link between gauge and geometrical description of General Relativity, giving rise to the fundamental relation
\beq\label{23}
F_{\mu\nu ab} = (R_{\mu\nu \a\be} + F_{\mu\nu \a\be})e^\a_a e^\be_b
\eeq
$R_{\mu\nu \a\be}$ being the Riemann tensor constructed with the metric $g_{\mu\nu}$ and $F_{\mu\nu \a\be}$ given by the relation
$$
F_{\mu\nu \a\be}  = \nabla_{[\mu}\bar A_{\nu]\a\be} + \bar A_{[\mu|\a\g|}A_{\nu] \ \be}^{\ \ \g}
$$
Using \eqref{22} and \eqref{23} equations \eqref{19}, \eqref{20} and \eqref{21} can alternatively be written as
\begin{align}
&e(0)^a_\mu = D_\mu\vf^a  \label{24}
\\[1ex]
& G_\mu^\nu + (F^\nu_\mu -\frac12\de^\nu_\mu F) -\chi\, T^\nu_\nu =0
\label{25}
\\[1ex]
&M^{\nu a\mu c}\bar A_{\mu\ c}^{\ b} + M^{\mu c\nu b} \bar A_{\mu c}^{\ \ a}\label{26} = 0
\end{align}
$G_\mu^\nu$ being  the Einstein tensor.

Using standard techniques to deal with \eqref{24}, \eqref{25} and \eqref{26} we arrive to the final result
\begin{align}
& e(0)_\mu^a = D_\mu\vf^a \label{27}
\\[1ex]
&A_{\mu ab} = e^\a_a\nabla_\mu e_{\a b} \label{28}
\\[1ex]
&G_\mu^\nu = \chi\, T_\mu^\nu \,.\label{29}
\end{align}

Now we have the elements to establish the geometrical properties of the background minimal gravity. At this purpose we take the $\la$--zero order term in equation \eqref{28}. Using then \eqref{28} and \eqref{27} we obtain 
$$
e(0)^a_\mu = \partial_\mu\vf^a + A(0)_{\mu\ b}^{\ a}\vf^b = \partial_\mu\vf^a + e(0)^{\a a}\nabla^{(0)}_\mu e(0)_{\a b}\vf^b
$$
where the covariant derivative $\nabla^{(0)}_\mu$ contains the zero order metric tensor $g(0)_{\mu\nu} = \eta ^{ab} e(0)_{\mu a}e(0)_{\nu b}$. The last can alternatively be written as
\beq\label{30}
e(0)_\mu^a = \partial_\mu\vf^a + e(0)^{\a a}\nabla^{(0)}_\mu\big[e(0)_{\a b}\vf^b\big] - e^{\a a}(0) e_{\a b}(0)\partial_\mu\vf^b\,.
\eeq
Taking now into account the definition of $e(0)_{\a b}$ we have 
$$
e(0)_{\a b}\vf^b = \big[\partial_\a\vf_b + A(0)_{\a bc}\vf^c\big]\vf^b = \partial_\a\s
$$
where $\s$ is a scalar field of the form
$$
\s = \frac12 \eta_{ab}\vf^a\vf^b\,;
$$
substituting this in \eqref{30} we get
$$
e(0)_\mu^a = e(0)^{\a a} \nabla^{(0)}_\mu \nabla^{(0)}_\a \s
$$
Thus multiplying by $e(0)_{\nu a}$ we finally deduce the value of the ``minimal" background metric tensor which becomes 
\beq\label{31}
g(0)_{\mu\nu} =\nabla^{(0)}_\mu\nabla^{(0)}_\nu\s\,.
\eeq
Curiously to reach  a result from the last is very easy. In fact, taking the trace in \eqref{31} one deduces $4= \nabla^{(0)\mu}\nabla^{(0)}_\mu\s \equiv \square^{(0)}\s$; then combining this one with \eqref{31} the condition
\beq\label{32}
\nabla^{(0)}_\mu \nabla^{(0)}_\nu \s = \frac14 g_{\mu\nu}\,\square^{(0)}\s
\eeq
is readily obtained.

As a general result the existence of solution for an equation of the form
$$
\nabla_\mu \nabla_\nu\, \s = \frac1{d}\, g_{\mu\nu}\,\square\s\,,
$$
$d$ being the dimension of the space, implies the maximally symmetric cha\-racter of the  the space, i.e.
$$
R_{\la\mu\nu\rho} = \frac{R}{d(d-1)}\big(g_{\la\nu}g_{\mu\rho} -g_{\la\rho} g_{\mu\nu}\big)\,.
$$

From these results and taking into account the Einstein field equations we conclude that in the absence of a energy--momentum term (pure gravity) the minimal background metric is, as expected, Minkowskian. The same holds when a traceless energy--momentum tensor is involved, this is for instance the case for a Maxwell--Einstein system. Nevertheless the occurrence of a cosmological constant will exclude the euclidean solution which should be substituted by a maximally symmetric background. We would like to point out that, assuming in \eqref{31} the flat solution, we directly obtain the metric $\eta_{ab}$ when we adopt as coordinates the Minkowskian ones $\vf^a$.

The formal structure of the general metric tensor can be derive from \eqref{16}. In fact, raising and lowering the Lorentz indexes we can rewrite it as
$$
e_{\mu a} = e(0)_\mu^b\big(\eta_{ba} + \la \Gamma^T_{ba}\big)\,,
$$
the remaining internal Lorentz symmetry in the latin indexes provide us with six additional conditions that can be exploited to render $ \Gamma^T_{ba}$ symme\-tric. Having this in mind we can write the general metric tensor as
\beq\label{33}
g_{\mu\nu} = g(0)_{\mu\nu} + 2\la \Gamma^T_{\mu\nu} + \la^2g(0)^{\rho\s}\Gamma^T_{\mu\rho}\Gamma^T_{\nu\s}
\eeq
where now $\Gamma^T_{\mu\nu}$ is symmetric and $g(0)_{\mu\nu}$ is the minimal background metric, we notice that doing this we have used the minimal tetrad $e(0)_\mu^a$ to change the nature of the indexes. We point out the resemblance between \eqref{33} and the usual weak field approximation. However it must be emphasized that \eqref{33} is an exact expression containing besides a second order term.

Thus schematically gravity can be described by the Einstein field equations with a metric given by \eqref{33} where $g(0)_{\mu\nu}$, that can be computed from \eqref{27} and \eqref{28}, is defined at $\la$--zero order in  a maximally symmetric space or even in some cases in an euclidean one as mentioned above.

\section{Final comments}

The present paper is mainly devoted to put in evidence the dynamical nature and internal structure of the tetrads in a Poincar\'e non linear gauge approach to General Relativity. In this treatment the dynamical objects are connections as in any other local field theory. However the very distinctive character of the translations requires the inclusion, as internal degrees of freedom, of Goldstone bosons isomorphic to Minkowskian coordinates. The introduction of a parameter $\la$ ``weighting" the pure translational connection gives support to a perturbative expansion. The validity of this expansion or, in other words, the relative smallness of $\la$ is, strictly speaking, a phenomenological hypothesis guaranteed nevertheless by the zero order result for the ``minimal" tetrad $e(0)_\mu^a$. Being a geometrical tool suitable to change holonomic and non holonomic coordinates between them, it can be colloquially described as ``gravity without gravity", notice the resemblance with the weak field expansion, we recall however that in our approach the covariance of the procedure is maintained. The prize to be pay is, of course, the necessity to calculate  $e(0)_\mu^a$ or $g(0)_{\mu\nu}$ order by order with the help of \eqref{27} and \eqref{28}.

As we have mentioned the origin and nature of $\la$ is in our treatment a reasonable phenomenological hypothesis, nevertheless, on very general grounds, it probably roughly describes deep properties of the intimate structure of the space--time, especially relevant, as we have point it out when translations are concerned.

There remains notwithstanding many open questions that  allowing us to consider besides this paper in some sense  as a proposal. To select some features to study in depth we mention, in the first place, the formal treatment as a symmetry of the translations when a critical length
is present.

The occurrence of an energy--momentum contribution with trace different from zero, cosmological constant or mass terms, for instance, implies a non euclidean background metric. This means that  our colloquial mention to a ``gravity without gravity" is at least to be taken very carefully.

We have been concerned in this work with a conservative scenario for gra\-vity including only a source for an energy--momentum contribution. The possible presence of a torsion source, as it happens, for instance when fermions are present, introduce interesting features without modifying the general scheme but increasing the complexity of the theory with additional torsion like terms.

\newpage
\section*{Appendix A: Parametrization of non linear\\ realizations}

The construction of a non linear realization of a group start from the choice of a suitable parametrization of the quotient space $G/H$ which, in practice, implies the use of a parametrization of the group space itself. We summarize in this appendix of the more commonly used techniques based on the well known properties of the exponentials.

Let be $G$ the whole group with $n$ parameters and $H$ a continuous subgroup of $G$. Let us call $\VEC{J}_i$ the generators of the subgroup 
algebra and $\VEC{A}_i$ the remaining ones of  $G$ not belonging to $H$.

Being $G$ a Lie group an arbitrary element $g\!\in\! G$  can  always be written uniquely in the form
$$
g = e^{\xi\VEC{A}}e^{u\VEC{J}}
$$
where $\xi \VEC{A}=\xi^i\VEC{A}_i$ and $u\VEC{J} = u^i\VEC{J}_i$

Now the action of a given element $g_0\!\in \!G$ over the elements of the quotient space $G/H$ is given by
$$
g_0\, e^{\xi\VEC{A}} = e^{\xi'\VEC{A}}e^{u'\VEC{J}}
$$
where $\xi'=\xi'(\xi,g_0)$ and $u'=u'(\xi,g_0)$

Let now be $h\!\in\! H$ and $\psi \to \rho(h)\psi$ a linear representation of $H$. It is evident that in this hypothesis the total action of $g_0$ is defined over $\xi$ and $\psi$ according to $\xi\to\xi'$ and $\psi\to\rho(e^{u'\VEC{J}})\psi$ which formally gives the non linear realization of $G$.

To verify the group law we consider an other element, say $g_1$, acting on $e^{\xi'\VEC{A}}$, i.e.
$$
g_1\, e^{\xi'\VEC{A}} = e^{\xi''\VEC{A}}e^{u''\VEC{J}}
$$
then
$$
g_1g_0\, e^{\xi\VEC{A}} = e^{\xi''\VEC{A}}e^{u'''\VEC{J}}
$$
where
$$
e^{u'''\VEC{J}} = e^{u''\VEC{J}}e^{u'\VEC{J}}\,.
$$
Being $\rho$ a representation it follows
$$
\rho(e^{u'''\VEC{J}}) = \rho(e^{u''\VEC{J}})\rho(e^{u'\VEC{J}})\,.
$$

Summarizing a non linear realization is, a few words, a technique that allows us to represent the action of a Lie group  over the linear representation of a given subgroup. The space of this realization is completed with the addition of the coset parameters which in an exponential parametrization strongly resembles the role of the Goldstone bosons.

\newpage
\section*{Appendix B: The Hausdorff--Baker--Campbell relations}

To present this paper as self--consistent as possible, we include the well known relations
$$
e^{-\VEC{A}} \VEC{B} e^{\VEC{A}} = \VEC{B} - 
\big[\VEC{A},\VEC{B}\big] +\frac1{2!} \Big[\VEC{A},\big[\VEC{A},\VEC{B}\big]\Big] -\frac1{3!} \Big[\VEC{A},\Big[\VEC{A},\big[\VEC{A},\VEC{B}\big]\Big]\Big] + \cdots
$$

$$
e^{-\VEC{A}}\, de^{\VEC{A}} = d\VEC{A} - 
\frac1{2!} \big[\VEC{A},d\VEC{A}\big]     +\frac1{3!}\Big[\VEC{A},\big[\VEC{A},d\VEC{A}\big]\Big] + \cdots
$$

$$
e^{\VEC{A}+\de \VEC{A}} = e^{\VEC{A}} +\de e^{\VEC{A}} +O\Big[(\de \VEC{A}^2)\Big] = e^{\VEC{A}}\big(\VEC{I} +e^{-\VEC{A}}\de e^{\VEC{A}}\big)
$$
useful to check calculations.

\end{document}